\documentclass{PoS}
\usepackage{natbib}

\title{Recent Advances in Cosmological Hydrogen Reionization}

\ShortTitle{Reionization}

\author{\speaker{Kristian Finlator}
        \thanks{I am grateful to the University of Texas Department of
	Astronomy Board of Visitors for supporting the 2011 
	Frank N.\ Bash Symposium.  I also thank NASA for support
	through Hubble Fellowship grant HF-51254.01 awarded by the
	Space Telescope Science Institute, which is operated by the 
	Association of Universities for Research in Astronomy, Inc., 
	for NASA, under contract NAS 5-26555.}\\
        UC Santa Barbara\\
        E-mail: \email{finlator@physics.ucsb.edu}}

\abstract{I discuss recent advances in the study of hydrogen reionization, focusing on
progress that was achieved during the years 2010--2011.  First, I discuss recent
measurements of the progress of reionization.  Next, I discuss recent observational
constraints on the nature and abundance of the dominant ionizing sources.  Finally,
I discuss recent progress in modeling reionization.  This review is written for
an audience of astronomers who do not specialize in the high-redshift Universe.}

\FullConference{Frank N. Bash Symposium New Horizons In Astronomy,\\
		October 9-11, 2011\\
		Austin Texas}

\begin{document}

\newcommand{\pasa}{Publications of the Astronomical Society of Australia}
\newcommand{\na}{NewA}
\newcommand{\aj}{AJ}
\newcommand{\mnras}{MNRAS}
\newcommand{\nat}{Nature}
\newcommand{\apss}{Ap\&SS}
\newcommand{\apj}{ApJ}
\newcommand{\apjl}{ApJ}
\newcommand{\apjs}{ApJS}
\newcommand{\aap}{A\&A}
\newcommand{\pasj}{PASJ}
\newcommand{\physrep}{Phys.~Rep.}
\newcommand{\araa}{ARA\&A}
\newcommand{\nar}{New A Rev.}
\newcommand{\ssr}{Space Science Reviews}

\newcommand{\kmsmpc}{\kms\;{\rm\ Mpc}^{-1}}
\newcommand{\lya}{Ly$\alpha$}
\newcommand{\hkpc}{h^{-1}{\rm\ kpc}}
\newcommand{\hmpc}{h^{-1}{\rm\ Mpc}}
\newcommand{\lcdm}{$\Lambda$CDM}
\newcommand{\kms}{\;{\rm km}\,{\rm s}^{-1}}

\newcommand{\mstar}{$M_{\star}$ }
\newcommand{\ud}{\mbox{\ d}}
\newcommand{\msun}{\mbox{M}_\odot}
\newcommand{\zsun}{\mbox{Z}_\odot}
\newcommand{\lgmstar}{\log(M_*/\msun)}  
\newcommand{\hinv}{h^{-1}}
\newcommand{\ebv} {$E(B-V)$}
\newcommand{\smyr} {\msun \mbox{ yr}^{-1}} 

\newcommand{\hi}{\hbox{H\,{\sc I}}}
\newcommand{\hii}{\hbox{H\,{\sc II}}}
\newcommand{\chii}{C_{\mathrm{H\,II}}}
\newcommand{\cb}{C_{\mathrm{b}}}
\newcommand{\fesc}{f_{\mathrm{esc}}}
\newcommand{\ncells}{N_{\mathrm{cells}}}
\newcommand{\tvir}{T_{\mathrm{vir}}}
\newcommand{\nhi}{n_{\mathrm{H\,I}}}
\newcommand{\nh}{n_{\mathrm{H}}}
\newcommand{\nhe}{n_{\mathrm{He}}}
\newcommand{\nhii}{n_{\mathrm{H\,II}}}
\newcommand{\nheii}{n_{\mathrm{He\,II}}}
\newcommand{\heii}{\mathrm{He\,II}}
\newcommand{\heiii}{\mathrm{He\,III}}
\newcommand{\ngrid}{n_{\mathrm{grid}}}
\newcommand{\tes}{\tau_{\mathrm{es}}}
\newcommand{\zreion}{z_{\mathrm{reion}}}
\newcommand{\xmxv}{x_{\mathrm{M}}/x_{\mathrm{V}}}
\newcommand{\xhim}{x_{\mathrm{H\,I,M}}}
\newcommand{\xhi}{x_{\mathrm{H\,I}}}
\newcommand{\xhiv}{x_{\mathrm{H\,I,V}}}
\newcommand{\xhiiv}{x_{\mathrm{H\,II,V}}}
\newcommand{\xhiim}{x_{\mathrm{H\,II,M}}}
\newcommand{\ngnb}{N_{\gamma}/N_b}
\newcommand{\lmfp}{\lambda_{\mathrm{MFP}}}
\newcommand{\dxr}{\Delta x_{\mathrm{R}}}
\newcommand{\dthr}{\Delta_{\mathrm{thr}}}
\newcommand{\dte}{\Delta t_{\mathrm{e}}}
\newcommand{\guvb}{\Gamma_{-12}}
\newcommand{\fcold}{f_{\mbox{\tiny cold}}}
\newcommand{\fbar}{f_{\mbox{\tiny bar}}}
\newcommand{\mh}{M_h}
\newcommand{\Cgas}{C_{\mbox{gas}}}
\newcommand{\rhodotstar}{\dot{\rho}_*}

\section{Introduction}
The Reionization Epoch consists broadly of the time interval between the 
moment when photons and matter decoupled, at a redshift $z$ of 
1088~\citep{sper03}, and the moment when enough ionizing sources had
formed to ionize the hydrogen and helium in the intergalactic medium (IGM), 
or $z\sim3$~\citep{shu10}.  This was the most active
period in the Universe's history: Galaxies were forming stars up to ten 
times as rapidly as they are today even as they expelled comparable amounts 
of gas back into their environments.  The supermassive black holes that are 
largely dormant today were each swallowing a dwarf galaxy's worth of gas 
over just a few billion years and glowing brightly enough to heat the IGM.  
Reionization converted the IGM into a plasma, suppressed the growth of 
dwarf galaxies, and altered the conditions under which subsequent 
generations of galaxies grew.  Its pivotal position at the outset of the 
story of structure formation rendered it one of the central science goals in 
the 2010 Decadal Survey~\citep{a2010}.

The Reionization Epoch is generally divided in two: During the hydrogen
reionization epoch, hydrogen was ionized and helium was singly ionized.  
During the helium reionization epoch, helium was further ionized from $\heii$ 
to $\heiii$.  Two considerations justify this division.  First, most of the 
photons that brought about hydrogen reionization were not energetic enough 
to remove the second electron from a helium atom.  At the same time, the 
cross section for hydrogen atoms to absorb photons that are energetic 
enough to ionize a $\heii$ atom is small.  Hence the dominant photons 
were different.  Second, hydrogen-ionizing photons were probably generated 
by massive stars in young galaxies while $\heii$-ionizing photons were 
predominantly generated by quasars.  Quasars did not grow abundant 
enough to participate in reionization until hydrogen reionization was 
largely complete~\citep{sha87}.  Hence the sources were also different.

This review focuses on hydrogen reionization (from now on, ``reionization").  
I shall discuss insights that have been won during 2010--2011, with 
reference to earlier results whenever necessary for context.  This field 
is routinely reviewed.  For a more technical introduction on early 
structure formation, see the classic discussions by Barkana and 
Loeb~\citep{bar01,loe01}.  For overviews of the first galaxies and their
role in reionization, see~\citep{cia05} and~\citep{bro11}.

I divide my discussion into three parts.  In Section~\ref{sec:xaxis},
I discuss progress in measuring the history of reionization.  When 
did reionization begin ? How long did it take? In Section~\ref{sec:sources}, 
I discuss progress in identifying and understanding the ionizing sources 
that drove reionization.  What were they? How many were there? Did the 
progress of reionization leave any observable signatures on their 
properties?  In Section~\ref{sec:models}, I discuss progress in modeling 
reionization.

\section{Measuring the Progress of Reionization}\label{sec:xaxis}
It is currently believed that reionization was driven by ionizing photons 
from dwarf galaxies.  If true, then its history can be divided into three 
stages~\citep{gne00}.  During the first stage, galaxies create enough 
ionizing photons to reionize their immediate surroundings, but the 
ionized regions remain small compared to the typical distance between 
galaxies.  During the second stage, ionized regions begin overlapping with 
each other and the typical distance traveled by ionizing photons grows 
rapidly.  During the final stage, the Universe is largely ionized and 
its residual opacity is dominated by slowly-shrinking ``islands" of 
moderately overdense gas.  In this Section, I discuss efforts to 
reconstruct when the Universe went through these stages.

\subsection{Lyman-$\alpha$ absorption}
The classic indication that reionization has already completed is the 
absence of a ``Gunn-Peterson trough" from the spectra of 
high-redshift quasars.  Intuitively, the cross-section for neutral 
hydrogen atoms to absorb photons that boost their electrons from the 
ground state (n=1) to the first excited state (n=2) is very large.  
Any photon that is in resonance with this transition (\lya) while 
traversing a region with neutral hydrogen will probably be 
absorbed.~~\cite{gun65} showed that, if the volume-averaged neutral 
hydrogen fraction $\xhi$ is $\sim1$, then the optical depth to 
scattering is $\sim10^4$.  Put differently, as UV photons redshift 
through resonance with \lya, they will on average be scattered unless 
fewer than 1 in $10^4$ hydrogen atoms is neutral.  In the 
post-reionization Universe, this phenomenon creates a ``forest" of 
absorption lines, where the amount of absorption at a rest-frame 
wavelength in the range  912-1216 \AA~traces the amount of neutral 
gas in the Universe at the moment when those photons redshifted 
through resonance with \lya.  In a reasonably neutral Universe, 
the forest converts into complete absorption, and is called the 
Gunn-Peterson trough.

The first reported Gunn-Peterson trough was identified in the spectrum 
of a quasar at $z=6.28$~\cite{bec01}, leading to speculation that 
reionization may have ended around this redshift.  In practice, 
however, the detection of a Gunn-Peterson trough only indicates that 
$\xhi>10^{-3}$~\citep{fan02}, which is still highly 
ionized.~~\cite{fan06} followed up on this by applying several 
complementary analyses to 19 quasars, tracing the evolution of 
$\xhi$ from $5\rightarrow6$.  
They found that $\xhi$ evolves rapidly during this interval and inferred
that the IGM is
mostly neutral at $z\geq6$.  In detail, their analyses all assumed 
a spatially homogeneous ionizing background, which is clearly an 
approximation near the end of reionization.  Relaxing this assumption 
leads to the more conservative limit $\xhi(z=6.1)<0.8$~\citep{mcg11}.

A complementary constraint on $\xhi$ comes from modeling
the ``proximity zones" of high-redshift quasars.  A proximity zone 
is a region centered on a quasar that is more ionized than the 
ambient IGM owing to the quasar's extra ionizing photons.  Its size, 
which can be measured and is typically a few Megaparsecs, is sensitive 
to (1) the quasar's ionizing luminosity; (2) the neutral hydrogen fraction 
before the quasar ``turned on"; and (3) the amount of time that the 
quasar has been active.  The quasar's ionizing luminosity can be 
constrained observationally, so if the duration of its active phase 
can be derived through other means then $\xhi$ can be inferred using a 
straightforward Str\"omgren Sphere calculation~\cite{sha87}.

Recently,~\cite{mor11} and~\cite{bol11} applied this
approach to a newly-discovered quasar at $z=7.085$ called 
ULAS J1120+0641.  Its proximity zone is unusually small in comparison 
to other high-redshift quasars, and it is the smallest when its radius
is scaled to the quasar's luminosity.  This is consistent with the IGM 
being more neutral than at lower redshifts.  Of course, it is also 
possible that the proximity zone is small because the quasar just 
``turned on".  Fortunately, the UKIDSS will soon yield 
more quasars at similar redshifts, enabling independent
measurements of $\xhi$~\citep{hew06}.

It is also possible to infer $\xhi$ from stronger \lya~absorbers.  If 
a region's neutral hydrogen column density exceeds $10^{20.3}$cm$^{-2}$ 
(and it will if the IGM is neutral), then \lya~absorption enters the 
damped portion of the curve of growth.  In this case, the red damping
wing's shape constrains $\xhi$.~~\cite{tot06} used the absence of a 
damping wing in the afterglow spectrum of a gamma-ray burst to infer 
that $\xhi(z=6.3) < 0.17$.   By contrast,~\cite{bol11} find evidence 
for this feature in the spectrum of ULAS J1120+0641, and they infer 
from it that $\xhi(z=7.085)>0.1$.  Of course, it is also possible 
that the surrounding IGM is highly ionized but a neutral parcel of 
gas (such as a satellite galaxy) happens to sit just in front of 
the quasar.  This possibility again indicates the need for a larger 
sample of reionization-epoch quasars.

\subsection{Redshifted hyperfine emission from neutral hydrogen}
A hydrogen atom has slightly lower energy when the spins of its 
proton and electron are antiparallel than when they are parallel,
and it can transition between these energy states by emitting or
absorbing a photon of wavelength 21.1 cm.  Early in the 
reionization epoch, when the IGM had been warmed by quasars and 
stars but not yet ionized, the entire Universe was glowing brightly 
in this transition.  Many of those photons were never re-absorbed, 
leading to a background that should be visible today at a frequency 
of 1420.4 Mhz$/(1+z)\approx140$ Mhz.  In fact, it should be 
possible to image the sky at a range of frequencies around 140 Mhz 
and document the progress of reionization~\citep{toz00} either by 
directly imaging brighter neutral and fainter ionized regions or 
through statistical approaches.  In practice, the observation is 
challenging because the expected flux is five orders of magnitude 
weaker than foreground sources.  Nonetheless, many low 
frequency radio experiments are under development that will use 
novel calibration techniques to isolate this signal and map out 
the history of reionization~\citep{mor10}.

Recently,~\cite{bow10} used the simpler approach of measuring the
all-sky spectrum of redshifted 21cm emission.  They inferred that, 
if reionization occured between $z=13$ and $z=6$, then its duration 
was longer than $\Delta z > 0.06$ with 95\% confidence.  Models generally 
predict that its duration was significantly longer than this, so it is 
not yet a strong constraint.  Nonetheless, it is the first result from 
what promises to be a rich field over the coming decades.

\subsection{The Cosmic Microwave Background}\label{ssec:tau}
Another approach to constraining the timing of reionization involves 
measuring the impact that Thomson scattering by free electrons had on 
the anisotropies in the cosmic microwave background (CMB).  This is 
quantified by $\tes$, which can be thought of as the probability that 
a given CMB photon scattered at some redshift below roughly 30.  
Measurements by the WMAP satellite indicate that $\tes=0.088\pm0.015$~\cite{kom11}.  
Of this, 0.050 originates from the portion of the photon's path between 
$z=7$ and $z=0$.  The remaining 0.035 is an integral constraint on the 
history of reionization.  It could indicate that reionization occurred 
over a short interval
(but longer than $\Delta z=0.06$!) around $z=10.6$, but it is equally
consistent with some scenarios in which reionization began at $z=14$
and completed around $z=6$ (see Figure 4 of~\citep{zah11b}).

Additional constraints come from much smaller-scale fluctuations in the
CMB.  The idea is that CMB photons can gain energy by inverse-Compton scattering 
off of energetic electrons.  If the scattering electrons reside within a 
galaxy cluster, then they create ``hot spots" in CMB maps; this is called 
the thermal Sunyaev-Zel'dovich (tSZ) effect.  If the scattering electrons 
exhibit bulk motions, then it is called the kinetic Sunyaev-Zel'dovich 
(kSZ) effect.  Some kSZ scattering occurs after reionization, but an additional
contribution is expected from ionized patches at $z>6$.  The kSZ signature 
constrains the duration of reionization because the distribution of ionized 
and neutral patches that a line of sight passes through depends on how 
long reionization lasted.  Since 2007, the South Pole Telescope (SPT) and 
the Atacama Cosmology Telescope (ACT) have been measuring small-scale 
fluctuations in the CMB that are sensitive to this effect.~~\cite{zah11b} 
recently combined measurements from the SPT and WMAP, finding with 95\% 
confidence that reionization had a duration of less than $\Delta z = 7.9$ 
and ended before $z=5.8$.  Their inferences will improve with more data 
and a better understanding of dusty star-forming galaxies at lower redshifts.
\subsection{Signatures of Reionization on Galaxies}\label{ssec:gals}
Star-forming galaxies irradiate their interstellar media (ISMs) with
ionizing photons from massive stars.  Two thirds of the resulting 
hydrogen recombinations produce \lya~photons.  Now, \lya~is the 
lowest-energy transition to the ground state of hydrogen.  This 
means that a \lya~photon in a neutral region will be repeatedly absorbed 
and re-emitted, performing a random walk until it either escapes or is 
destroyed.  The photon is ``destroyed" if an absorbing atom decays 
through two-photon emission.  It can also be destroyed through dust 
absorption or OIII Bowen fluorescence.  If the IGM is fully ionized, 
then any photons that escape their galaxy's ISM will no longer scatter.  
This is the case at $z<6$, where many galaxies show strong \lya~emission 
lines~\citep{nil09}.  If a galaxy emits \lya~photons into an IGM that 
is not fully ionized, however, then its \lya~photons will be scattered 
into a low surface-brightness halo.  This has several measureable 
consequences. 

{\bf The luminosity function of \lya~emitters should evolve.}
Galaxies that are selected to have bright \lya~emission lines are
called \lya~emitters (LAEs).  Models indicate that, at epochs during 
which the IGM is more than 50\% ionized, the observed abundance of LAEs 
should drop precipitously owing to IGM scattering~\citep{fur06a,mcq07}.  
Recently, three studies reported that the luminosity function (LF) of 
LAEs decreases from $z=5.7$ to $z=6.5$~\citep{hu10,ouc10,kas11}.  This 
evolution could owe to a 
partially-neutral IGM at $z=6.5$, but it could also owe to an overall 
decline in the abundance of galaxies.~~\cite{kas11} therefore verified 
that the distribution of UV-continuum fluxes from their LAEs at $z=5.7$ 
and $z=6.5$ are similar.  This conflicts with the view that the $z=6.5$
sample is fainter in \lya~owing to lower star formation rates and
suggests a partially-neutral IGM.  If so,~\citep{kas11} inferred that 
$\xhi$ increases from 0 to 0.3--0.4 from $z=5.7\rightarrow6.5$.  However,
the amount by which the LAE LF evolves differs between the different
studies, leading to uncertainty in the strength of the constraint on 
reionization.  Larger samples will eventually settle this question.

{\bf The \lya~escape fraction should evolve.}
A complementary way to identify IGM attenuation on \lya~emission 
lines is to identify galaxies at a range of redshifts using a 
selection technique that does not involve \lya, then explore the
evolution in their \lya~properties.  Samples that push into
the reionization epoch should have suppressed \lya~emission
for the same reasons as before.  Galaxies that are identified 
via the Lyman-dropout technique can be used in this 
way~\citep{sha11}.

Over the past two years, evidence has accumulated that the 
equivalent width of \lya~emission from such samples grows with 
increasing redshift to $z=6$ and then drops to higher 
redshifts~\citep{ouc10,ono11,pen11,sch12,sta11}.  This evolution 
is often referred to as an evolving ``escape fraction" of \lya, 
but its physical interpretation remains controversial.~~\cite{ono11}
present evidence that the \lya~escape fraction drops for faint 
objects but not for bright ones and interpret this as a constraint
on the topology of reionization, or the order in which regions of
different densities reionized.  This conflicts with the expectation
that reionization will suppress the abundance of bright \lya~sources 
by the same factor as faint ones~\citep{fur06a,mcq07}.  Therefore, 
differential evolution in the \lya~escape fraction could also be a 
signature of differential evolution in galaxies' ISMs.  If, for 
example, the gas fraction evolves rapidly for low-mass galaxies,
then they could simply destroy their \lya~photons more effectively 
at higher redshift.  Settling this question will require improved 
constraints on bright and faint galaxies.

{\bf The clustering of LAEs should increase.}
Galaxies are not distributed uniformly in space, and the degree
to which they cluster constrains the dominant growth processes.
Conceptually, observers quantify clustering through the enhanced
probability that a galaxy has a neighbor at a given distance over what 
would be expected if galaxies were distributed  uniformly.  If galaxies 
produced the photons that drove cosmological reionization,
then their clustering influences the history of reionization.  In 
particular, by the time that $\xhi$ has dropped to $\sim50\%$, the 
average ionized gas parcel sits within an ionized region that is of 
order 10 comoving Megaparsecs in radius and whose ionization is 
maintained by many hundreds of galaxies working together.  
A \lya~photon emitted within such a region redshifts out of resonance
with \lya~after traveling $\sim1$ Mpc~\citep{cen00}.  By the time it 
reaches the neutral IGM, it will no longer be able to scatter.  The 
implication is that LAEs ought to 
exhibit increased clustering as observations push into the
reionization epoch because \lya~photons from galaxies in crowded 
neighborhoods will travel farther before reaching the neutral 
IGM.~~\cite{ouc10} searched for this signal and found that LAEs' 
clustering properties at $z=6.6$ were indistinguishable from samples 
at lower redshifts, suggesting that the IGM was already more than
50\% ionized by $z=6.6$.

{\bf The mean shape of the \lya~line profile should evolve.}
As $\xhi$ increases, the tendency for the IGM to scatter the
\lya~line should evolve more strongly on the blue side of the
line than on the red side.  Consequently, the typical profile
of the \lya~line could evolve in redshift.~~\cite{ouc10}
compared the stacked line profiles of LAEs at $z=5.7$ and $z=6.6$,
finding at most $1\sigma$ evidence for evolving line profiles.
This is inconsistent with some models in which $\xhi$ increases
from 0 to $>0.5$ during this redshift interval.

In summary, some of the expected signatures of reionization on
the \lya~properties of galaxies have been
detected while others have not.  Unfortunately, the 
interpretation of these observations remains unclear.  A major 
source of uncertainty involves the role that bulk gas motions
could play in modulating \lya~line profiles.  For example, 
gas outflows can endow \lya~photons with a net redshift, making
it easier for them to avoid scattering off of the IGM once
they emerge from the galaxy.~~\cite{dij11} showed that this 
effect can allow more than half of all \lya~photons to avoid scattering 
even if $\xhi=0.6$.  More generally, a consensus as to the physical 
interpretation for \lya~line profiles does not yet exist owing to 
degeneracies between the processes that could affect them.  For this 
reason, observations of \lya~emission cannot by themselves 
constrain the history of reionization even though they yield a 
variety of useful and complementary clues.

\section{Measuring the Sources of Reionization}\label{sec:sources}
Reionization clearly happened.  Hence while some investigators
reconstruct its timing, others ask which ionizing sources could 
have been responsible.  The sources that can be constrained 
observationally are quasars and galaxies; I shall 
touch on more speculative sources in Section~\ref{sec:models}.

Quasars did not dominate reionization.  In particular, by measuring 
their abundance, estimating their total ionizing luminosity, and 
comparing it to the IGM recombination rate, many studies have shown 
that quasars could not have contributed more than half of all ionizing 
photons at $z>4$~\cite{sha87,gli11} and only 1--5\% at 
$z=6$~\cite{wil10}.  This leaves galaxies, which in turn raises two 
questions: (1) How many galaxies were there as a function of luminosity 
and redshift? and (2) How many ionizing photons did they emit into the
IGM?

{\bf Counting Galaxies}
By counting galaxies and making simple assumptions regarding their
ionizing luminositites, it is possible to show that the observed 
galaxies could not have kept the Universe ionized at $z=6$.  
What about fainter galaxies? The extrapolated abundance of faint 
galaxies is quantified by the slope of the power law that fits 
the LF's faint end.  Galaxies may drive reionization as long as 
this slope, $\alpha$, is steeper (more negative) than 
-1.6~\cite{yan04,rob10}.  A number of groups have recently used 
extraordinarily deep images taken by the Wide Field Camera 3 (WFC3) 
on board the Hubble Space Telescope (HST) to constrain the abundance 
of Lyman dropout galaxies out to 
$z=10$~\citep{fin10,bou11a,dun11,gon11,gra11,mcl11,oes12}.
A consensus is emerging that $\alpha$ is between -1.8 and -2 
for $z\geq6$.  In a complementary study,~\cite{hen12} measured the 
faint-end slope of the LAE LF  at $z=5.7$, finding $\alpha=-1.7$.  
Both results are consistent with the view that faint galaxies 
drove reionization.

A complementary way to constrain the activity in galaxies involves 
counting gamma-ray bursts (GRBs).  ``Long" gamma-ray bursts are 
associated with core-collapse supernovae, hence their abundance 
tracks the total star formation rate density.  This inference is 
subject to its own systematic uncertainties, but unlike 
galaxy-counting studies it does not require a correction for the
contribution from faint galaxies.  Recent work indicates 
that the GRB abundance implies enough star formation in galaxies
to complete reionization by $z\approx8$~\citep{kis09,wyi10,rob12}.

{\bf Inferring Their Ionizing Luminosities}
How many ionizing photons did each galaxy emit into the IGM? This
depends on the metallicity and age of its stars as well as the 
amount of dust in its ISM.  Each of these properties also impacts 
the ultraviolet (UV) continuum ($\lambda\approx$912-3000 \AA) of 
a galaxy's spectrum.  Broadly, galaxies with young stars,
low metallicities, and low amounts of dust have bluer UV 
continua and higher ionizing luminosities.  Hence constraining the 
UV continua of reionization-epoch galaxies is a way to learn 
about their ionizing luminosities.  

Recent studies have used WFC3 observations to show that fainter
reionization-epoch galaxies are bluer~\citep{bou11b,fink11}.
They interpret this trend as evidence that fainter galaxies have 
less dusty ISMs.  If so, then dust was nearly absent from 
the faint galaxies that probably dominated reionization.  The UV
continuum luminosity is then a more direct tracer of the ionizing 
luminosity than at lower redshifts, where the UV continuum is 
invariably modulated by dust extinction.  Moreover, a larger 
fraction of ionizing photons probably escape into the IGM.  Both 
of these
studies also find that the UV continuum slopes lie within the range 
expected for moderately metal-enriched stellar populations.  In 
other words, observations do not yet provide evidence for the
long-sought zero-metallicity stars, which are expected to show 
very blue UV continua~\citep{sch02}.

What fraction of ionizing photons escape into the IGM and participate
in reionization? Clearly, this question is impossible to answer 
directly because ionizing photons emitted by reionization-epoch 
galaxies were absorbed by the IGM.  However, it is possible to 
constrain the ionizing escape fraction, $\fesc$, at lower redshifts.
~~\cite{sia10} and~\cite{nes11} recently used direct detections 
of ioinizing photons from galaxies at $z\sim3$ to show that the 
ratio of escaping ionizing continuum flux $F_{\mathrm{LyC}}$ 
to 1500 \AA~flux $F_{1500}$ increases both with increasing redshift 
and decreasing luminosity.  This suggests that $\fesc$ increases to 
high redshift and faint luminosity.  Remarkably, their measurements 
also favor higher ratios of $F_{\mathrm{LyC}}/F_{1500}$ than is
expected of standard stellar population sythesis models.  Both
findings enhance the role of faint galaxies in driving reionization.

\section{Modeling Reionization}\label{sec:models}

\subsection{Sources of Reionization}
Although it is widely believed that galaxies and quasars dominated
reionization, they need not have acted alone.  The theoretical 
search for plausible additional ionizing sources reads somewhat like a 
murder mystery in which the crime occurs in a crowded room.~~\cite{joh11} 
recently argued that supernova remnants could have emitted up to 
10\% as many ionizing photons into the IGM as massive stars.  The 
idea is that shocks produce a harder ionizing continuum than hot 
stars.  Their more energetic photons experience a longer mean free 
path, hence their contribution to the photons that escape into the 
IGM could be significant.

Shocks from structure formation have also been 
forwarded~\cite{min04,dop11}.  Here,
the idea is that gas that accretes into relatively massive halos 
($>3\times10^{10}\msun$ for~\cite{dop11} and $10^{11}-10^{12}\msun$
for~\cite{min04}) shocks as it accretes and virializes, converting
gravitational potential energy into ionizing photons.~~\cite{wyi11} 
recently showed that, even if all baryons that fall into massive 
halos convert their gravitational potential energy into photons at 
the hydrogen ionization edge, they only produce 1 ionizing photon 
per three hydrogen atoms by $z=6$.  This is far less than the likely 
contribution from galaxies.  Nonetheless, these inquiries suggest 
that shocks may have impacted the IGM's thermal history, the topology
of hydrogen reionization, and played a role in $\heii$ reionization.

Star formation in the progenitors of globular clusters (GCs) has also
been proposed~\cite{ric02}.  This is especially plausible if the total 
mass that formed in the progenitors of present-day GCs was 
$\sim10\times$ as large as their current stellar masses and if 
$\fesc\sim1$.  This idea has recently gained momentum from new 
observations of the metal abundances of GCs and the identification of 
candidate second-generation stars in the Milky Way's stellar halo.  By 
using these observations to constrain the stellar initial mass
function of GCs as well as the mass function of GCs 
themselves,~\cite{sch11} have shown that the contribution of GCs to 
reionization could have been quite large.  If true, then the
progenitors of massive GCs could have had masses of $10^6\msun$
and will be visible to JWST.

More exotic sources of ionizing photons such as decaying dark
matter particles and evaporating primordial black have not received 
much attention in recent years.  We refer the reader to Section 
3.4.4 of~\citet{fur06b} and Section 1 of~\citet{cho06} for an 
introduction to these alternatives.

\subsection{The Ionizing Escape Fraction from Galaxies}
The fraction of ionizing photons that escaped into the IGM, $\fesc$, 
is a crucial parameter that is regularly studied 
theoretically. \cite{fer11} review both observational and theoretical 
inquiries 
into $\fesc$; interested readers may refer to their introduction for 
an overview.  Broadly, analytical and numerical models predict values 
that span the full range from 0--1.  In order to understand this 
apparent lack of consensus,~\cite{fer11} use a simple model of stars 
embedded in a disk to explore how the properties of a galaxy's ISM 
and its stellar population modulate $\fesc$.  They show that 
galaxies that deposit a large fraction of their ISM into a small 
number of high-density clumps possess a large $\fesc$.
They also show that more intensely star-forming galaxies possess 
higher $\fesc$ partly because fewer baryons remain in the 
ISM and partly because the number of optically-thin sightlines 
through the ISM increases.  They then attribute the wide range 
spanned by theoretical studies primarily to differences in the 
assumed gas masses and ISM topologies.  The implication is
that the predicted $\fesc$ will remain uncertain until the ISMs 
of high-redshift galaxies are understood.

\subsection{The IGM Recombination Rate}
In Section~\ref{sec:sources}, I discussed efforts to measure the
rate at which new ionizing photons were emitted into the IGM.  
Closing the loop requires us to know how many photons were required 
to achieve reionization, which in turn depends on the IGM's mean
recombination rate.  At early times ($z\geq20$), the Universe was 
relatively homogeneous and the recombination rate per volume in an 
ionized region was well-approximated by $\alpha\langle\nh\rangle^2$, 
where $\alpha$ is the hydrogen recombination coefficient at a
characteristic temperature such as $10^4$K and $\langle\nh\rangle$ 
is the mean hydrogen number density.  By $z=6$, however, the IGM
was inhomogeneous.  Correcting for this involves multiplying 
the above rate by the clumping factor $C$, which depends on the 
probability distribution function of baryon densities, the topology 
of reionization---that is, which regions are ionized and which ones 
remain neutral---and the temperature of the ionized gas.~~\cite{paw09} 
used realistic density fields culled from three-dimensional numerical 
simulations to predict that $C$ lies within the range 3--8 by $z=6$.  
More recently,~\cite{shu11} used a different numerical model to 
predict that $C=$2--4 by $z=6$.  Neither of these works included a 
self-consistent treatment of cosmological reionization, hence they 
were forced to make assumptions regarding the densities and 
temperatures of ionized regions.  Their simulations also subtended 
different cosmological volumes at different spatial resolutions, 
hence it is possible that the slightly lower value of~\cite{shu11} 
owes to differences in resolution and the assumed topology of 
reionization.  Broadly, however, both works prefer values $C$ that 
lie within the range of 3--10.  This range is low enough to allow
dwarf galaxies to drive reionization.

\subsection{One-Dimensional Models}
As we have seen, the history of reionization remains subject to 
many uncertainties such as the ionizing emissivity of stars and
quasars, $\fesc$, $C$, and the way in which these quantities evolve.  
A popular approach to understanding how these factors interact 
involves distilling reionization to a single equation that follows 
the growth of the ionized volume fraction with time.  In this 
``photon-counting" exercise (for example, Equation 64 
of~\citep{haa12}), the time rate of change of the neutral volume 
fraction is given by $-$(the ionizing emissivity per hydrogen 
atom from all sources) $+$(the recombination rate per hydrogen atom).  
The first term includes $\fesc$ and the second includes 
$C$.~~\cite{haa12} have shown that this formalism brings observations 
of the abundances of galaxies, quasars, IGM absorbers, and the optical 
depth to Thomson scattering (Section~\ref{ssec:tau}) into agreement 
with one another.  Importantly, they assume a relatively low value 
$C=3$ at $z=6$, and they assume that the luminosity-weighted mean 
$\fesc$ increases at earlier times such that it exceeds 50\% for 
$z>9.3$.  These insights serve both as valuable inputs to 
complementary theoretical efforts and as predictions for future 
observational campaigns that will constrain $\fesc$.

\subsection{Three-dimensional Models}
Theorists have been attempting to model reionization in three 
dimensions for over a decade.  A central goal is to model cosmic 
structure formation starting from $z\geq100$ with as few physical 
assumptions as possible while reproducing as many observables as 
possible.  I will divide the factors that models must consider into 
``agents", ``processes", and ``parameters".  Agents include density 
fluctuations, Lyman limit systems, galaxies, and quasars.  Processes 
include gas flows, the absorption of ionizing photons by an 
inhomogeneous IGM, recombinations, spectral hardening, and Jeans 
smoothing (Section~\ref{ssec:nbody}).  Emerging parameterizations 
that quantify these processes include the star formation efficiency 
as a function of dark matter halo mass; $\fesc$; the latent heat per 
IGM photoionization; and the IGM temperature, metallicity, ionization 
state, and recombination rate.  To date, the chief observables have 
consisted of $\tes$ (Section~\ref{ssec:tau}) and $\xhi(z)$.  In the 
near future, reionization models will begin confronting complementary 
observations of galaxies, the IGM metallicity, and the 
post-reionization Universe more generally.

The challenge of modeling these processes self-consistently is 
often summarized as follows: Properly 
sampling long-wavelength density fluctuations requires computational 
volumes to subtend at least 100$\hmpc$~\citep{bar04}.  Meanwhile, the 
lowest-mass dark matter halos that can form stars have a total mass of 
$\sim10^8\msun$; these must be resolved with $>100$ resolution elements. 
This implies a spatial dynamic range of $3\times10^5$.  State-of-the-art 
hydrodynamic simulations currently achieve one tenth of this range.
Folding in a treatment for radiation transport increases the 
computational expense by an additional 1--2 orders of magnitude.
This dynamic range can be achieved within individual regions in 
``zoom-in" simulations, but not throughout a representative 
cosmological volume.  In the face of these demands, progress is
made by omitting or treating approximately some subset of 
the relevant physical processes or scales.  Here I shall review 
recent efforts to this end.  For a more comprehensive overview of 
cosmological radiation transport, see the code comparison 
papers~\citep{ili06,ili09} or the review article by~\citet{tra09}.

\subsubsection{Dark Matter + Radiation Transport}\label{ssec:nbody}
Many studies use N-body (gravity-only) simulations to derive the cosmic 
density field, assume that gas follows matter, populate dark matter 
halos with ionizing sources following simple prescriptions, and then 
compute the inhomogeneous reionization history (see~\citep{tra09}
for a summary of these works).  The most recent of these calculations 
benefit from excellent spatial dynamic range, and they have 
simultaneously reproduced observations of $\tes$ and $\xhi$.  The 
tradeoff is that they do not treat hydrodynamics and must therefore 
make assumptions 
regarding the scaling between ionizing luminosity and halo mass, the 
impact of photoionization heating on the IGM clumping factor, and the 
response of low-mass systems to photoionization heating.  The latter
processes are called ``Jeans smoothing" and ``Jeans suppression" and are 
simple to understand: When mildly overdense regions are photoheated, 
they expand until their self-gravity balances the local gas pressure
(Jeans smoothing).  Dark matter halos with total masses below 
$5\times10^8\msun$ cannot accrete gas in such regions~\citep{fin11}, 
hence their star formation is quenched within a few dynamical times
(Jeans suppression).

Another approach is to derive the cosmological density field from 
precomputed hydrodynamic simulations, which can account (though 
not self-consistently) for Jeans smoothing by assuming a precomputed 
ionizing background.  Such studies have also confirmed that galaxies 
could have driven reionization despite the Jeans suppression of 
low-mass sources.  Interestingly, they fail to reproduce the 
post-reionization $\xhi$ and ionization rate.  This could owe to 
resolution limitations~\citep{fin09} or to observational 
systematics~\citep{aub10}.

In a pathbreaking work,~\cite{tra08} combined a high-resolution 
N-body calculation of the density field with a medium-resolution 
hydrodynamical calculation that directly models Jeans smoothing 
while omitting Jeans suppression.  They found that, following 
reionization, the IGM temperature and density are inversely related.
This is because overdense regions are reionized and photoheated first, 
and have therefore had the most time to cool by $z=6$.  Observational 
tests of predictions such as these will provide complementary 
constraints on the history of reionization.

\subsubsection{Semi-analytical Models}
Semi-analytical models (SAMs)~\citep[for example,][]{som01} 
treat the formation of dark matter halos with relatively few 
approximations, then model how galaxies grow within those halos by 
using analytic prescriptions for baryonic processes such as gas 
infall, outflows, and star formation.  By sidestepping the 
expense of discretizing the gas into parcels and integrating the
gas equations, they sacrifice some realism in exchange for expanded 
flexibility.  The governing physical parameters can, however, be 
tuned using observations and numerical simulations, improving 
their realism.

\citet[][and references therein]{ben06} adapted a SAM 
to compute reionization self-consistently by modeling how
ionizations and recombinations drive $\xhi(z)$.  Their approach 
accounts for Jeans suppression and includes a model for the clumping 
factor.  It treats radiative processes such as self-shielding and 
shadowing only approximately and does not predict the topology of 
reionization.  However, its ionizing emissivity can be tuned through 
observations of galaxy and quasar abundances in the post-reionization 
Universe, a major advantage.  By varying parameters such as $\fesc$
and the strength of star formation feedback, they showed that 
galaxies could give rise to a range of reionization histories 
including some that are consistent with observations.

State-of-the-art SAMs use a statistical approach to confront 
many complementary observations simultaneously, yielding 
joint constraints on processes that regulate galaxy 
evolution~\citep[for example,][]{lu11}.  Augmenting these models with 
treatments for inhomogeneous reionization will render them a powerful 
complement to numerical simulations.

\subsubsection{Semi-numerical Models}
If reionization was driven by galaxies, then the ionizing photons
experienced a short mean free path and, on large enough scales,
reionization was a local process.  Semi-numerical models build upon 
this idea: if the number of ionizing photons produced within a region 
exceeds the number of hydrogen atoms (corrected for recombinations), 
then the region is probably ionized.  These ideas were introduced 
within the context of an analytical model for 
reionization~\citep{fur04}, and have since been adapted successfully 
for three-dimensional volumes~\citep[for example,][]{zah05}.  
They reproduce the spatial distribution of neutral and ionized regions 
predicted by the more exact numerical methods on scales larger than 
$\sim1\hmpc$~\citep{zah11a}.  Given that they are inexpensive
computationally, they can be combined with a statistical formalism 
to invert observations of the IGM ionization state and constrain 
the history of reionization~\citep[for example,][]{zah11b}.  
They cannot treat processes that occur on spatial scales below which 
the assumption of local ionization does not apply such as spectral 
filtering, shadowing, and self-shielding within overdense regions.  

\subsubsection{Cosmological Radiation-Hydrodynamic Simulations}
Simulations that treat dark matter, gas, and the radiation field
self-consistently are the most realistic and 
computationally-intensive approach to modeling reionization.  
Their principal advantage is that they accurately model small-scale 
processes such as Jeans smoothing and Jeans suppression.  The tradeoff 
is that their computational expense prevents them from modeling 
volumes that are larger than $(\sim10\hmpc)^3$, hence they do not 
capture long-wavelength density fluctuations.

\citet{gne06} compared predictions from cosmological 
radiation hydrodynamic simulations with observations of the 
reionization-epoch \lya~forest and $\tes$.  They showed that 
high-resolution simulations subtending 4 and 8$\hmpc$ volumes 
yield converged predictions of the \lya~forest during the 
redshift range $5 < z < 6.2$ if the ionizing efficiency is 
adjusted to reproduce observations at $z=6\pm0.1$.  The
predicted $\xhi(z>6.1)$ and $\tes$ depend on the adopted 
mass resolution owing to star formation activity in systems whose 
mass lies near or below the resolution limit.  The predicted IGM 
transmission grows less resolution-convergent in the reionization 
epoch because it is dominated by rare optically-thin regions.  
This work showed that simulations can reproduce the evolution of 
several observables of the reionization-epoch IGM.  However, 
accounting for their sensitivity to resolution and volume 
limitations represents a formidable challenge.

In order to leverage the extensive observational constraints
available from the post-reionization epoch,~\citet{fin11} merged a
well-studied galaxy evolution model with a radiation transport
solver.  They then used this framework to explore the impact of 
outflows and Jeans smoothing on star formation and 
reionization.  They found that models combining strong outflows with 
an escape fraction of $\fesc=50\%$ simultaneously reproduces the 
observed galaxy abundance at $z=$6--8 while completing reionization
by $z=6$.  They also studied how outflows and an ionizing background 
couple, finding that outflows promote star formation in dwarf galaxies
by delaying reionization while suppressing it in more massive systems 
by coupling nonlinearly to the ionizing background~\citep{paw09}.  
Their preferred model overproduced the strength of the ionizing 
background at $z<6$ while underproducing $\tes$.  They interpreted 
this as evidence that $\fesc$ must evolve (consistent with~\citep{haa12}).

The problem of how to use radiation hydrodynamic simulations to improve
our understanding of reionization-epoch galaxies and the IGM is a
theoretical frontier owing to its computational expense.  Over the next
decade, efforts will be directed at two complementary goals.  First, 
models will be generalized in order to expand the range of observables 
against which they can be tested.  For example, artificial lines of sight 
through simulations will be compared with quasar absorption spectra in 
order to measure the reionization-epoch IGM's metallicity and temperature.  
Likewise, testing 
models against measurements of galaxies' colors and luminosities will 
constrain the abundance and activity in the putative ionizing population.  
These studies can be undertaken using existing simulations and will test 
current assumptions.  Relaxing those assumptions is the second goal, 
and doing so requires expanding the dynamic range.  Here there are two
ways forward.  The ``brute-force" approach involves chipping away at the
limitations of monolithic simulations through improved algorithms
and hardware.  The other approach uses high-resolution, small-volume
simulations to tune the assumptions that underlie cosmological 
simulations.  For example,~\citep{koh07} used small-volume simulations to
estimate $C$ while other authors are using high-resolution simulations 
to model the star formation rates, metallicities, and $\fesc$ from 
reionization-epoch dwarf galaxies~\citep{bro11}.  This approach 
will remain important until dynamic range limitations are overcome.

\section{Summary}
Over the past decade, CMB and galaxy observations have given rise to 
a consensus that galaxies could have driven hydrogen reionization, 
which in turn ended at some point before $z=6$.  While little else is 
clear at the moment, a wide variety of studies are whittling away at 
the dominant questions.  New observations of emission lines, the CMB, 
and redshifted 21 centimeter emission will soon measure the history 
of reionization.  Ground- and space-based
observations will locate and characterize the sources that caused it.  
On the theoretical side, ``brute force" campaigns are expanding the 
dynamic range of numerical simulations.  At the same time,
approximate techniques are under development that will synthesize 
observational and theoretical insights into models 
that have broad applicability.  All of this activity will
render the coming decade a very active one, and there is every 
reason to believe that it will yield a much clearer picture of what 
went on during the second half of the first billion years.

\renewcommand{\baselinestretch}{0.8}\normalsize
\bibliographystyle{abbrv}

\normalsize

\end{document}